\documentclass[]{interact}

\usepackage{epstopdf}
\usepackage[caption=false]{subfig}
\usepackage{color}

\usepackage[numbers,sort&compress]{natbib}
\bibpunct[, ]{[}{]}{,}{n}{,}{,}
\renewcommand\bibfont{\fontsize{10}{12}\selectfont}
\makeatletter
\def\NAT@def@citea{\def\@citea{\NAT@separator}}
\makeatother

\theoremstyle{plain}

\theoremstyle{definition}

\theoremstyle{remark}

\newcommand{\Langle}{\left\langle}
\newcommand{\Rangle}{\right\rangle}
\newcommand{\eq}[1]{Eq.~(\ref{#1})}

\begin{document}

\articletype{RESEARCH ARTICLE}

\title{Generating multi-wavelength phase screens for atmospheric wave optics simulations using fast Fourier transforms}

\author{
\name{M.~W. Hyde~IV\textsuperscript{a}\thanks{CONTACT M.~W. Hyde~IV. Email: mhyde@epsilonsystems.com}, M.~F.~Spencer\textsuperscript{b,c}, and M. Kalensky\textsuperscript{d}}
\affil{\textsuperscript{a}Epsilon C5I, Beavercreek, OH, 45431, USA\\ \textsuperscript{b}Department of Engineering Physics, Air Force Institute of Technology, Dayton, OH, 45433, USA\\ \textsuperscript{c}Joint Directed Energy Transition Office, Office of the Under Secretary of Defense for Research and Engineering, Washington, DC, USA\\  \textsuperscript{d}Integrated Engagement Systems Department, Naval Surface Warfare Center Dahlgren Division, Dahlgren, Virginia, USA}}

\maketitle

\begin{abstract}
We present a method to synthesize phase screens for multi-wavelength, atmospheric wave optics simulations using fast Fourier transforms.  We validate our work by comparing the theoretical, two-wavelength optical path length structure function to simulated results, and we find the two to be in excellent agreement.  Our phase screen synthesis method will be useful in two-wavelength adaptive optics simulations in strong or deep turbulence.        
\end{abstract}


\section{Introduction}
Techniques to quickly and accurately synthesize atmospheric phase screens for use in wave optics simulations has been an active area of research for over 30 years~\cite{10.1117/12.55712, Lane,Roggemann1996,Frehlich:00,Schmidt:10,Carbillet:10,Charnotskii:13,Charnotskii:13_3,Charnotskii:20}.  For much of this history, the primary goal has been to generate an accurate, discrete representation of a continuous turbulent path at a single wavelength.

Recently, there has been interest in two- or multi-wavelength adaptive optics (AO) systems for beam projection applications~\cite{Hyde:24,10529268}.  In two-wavelength AO, the atmosphere is sensed at one wavelength and corrected at another.  In astronomy, this is common practice and has been studied for decades~\cite{fugate1991measurement, Fugate:94,LGSA,Parenti:94,Wang:12, Lukin:79, Hogge:82,Holmes:83,Winocur:83,Devaney:08}.  However, in contrast to astronomical viewing, two-wavelength AO systems for beam projection generally operate in complementary conditions (i.e., small fields of view and long, nearly horizontal paths \cite{Kalensky:24}).  The latter causes scintillation which gives rise to phase discontinuities known as branch points~\cite{Fried:92,Fried:98}.  The effect of these phenomena on two-wavelength AO performance has yet to be fully quantified~\cite{Hyde:24}.

This, of course, motivates research into two-wavelength AO systems in strong turbulence conditions and, as a consequence, techniques to generate turbulence realizations with the proper statistics at two or more wavelengths.  Indeed, the two-wavelength phase covariance function $B_S$, first derived by Ishimaru~\cite{1140133}, is
\begin{equation}\label{eq:2}
\begin{split}
B_S\left(\boldsymbol{\rho},k_1,k_2\right) &= \Langle \phi\left(\boldsymbol{\rho}_1,k_1\right) \phi\left(\boldsymbol{\rho}_2,k_2\right)\Rangle = \iint_{-\infty}^{\infty} \Phi_S\left(\boldsymbol{\kappa},k_1,k_2\right) \exp\left(\text{j}\boldsymbol{\kappa}\cdot \boldsymbol{\rho}\right) \text{d}^2\kappa \hfill \\
\Phi_S\left({\kappa},k_1,k_2\right) &= \pi k_1 k_2 z \Phi_n\left(\kappa\right) \left\{ \operatorname{sinc}\left[\frac{z}{2}\left(\frac{1}{k_1}-\frac{1}{k_2}\right)\kappa^2\right] +\operatorname{sinc}\left[\frac{z}{2}\left(\frac{1}{ k_1}+\frac{1}{k_2}\right)\kappa^2\right]\right\}, \hfill
\end{split}
\end{equation}
where $\boldsymbol{\rho}=\boldsymbol{\rho}_1-\boldsymbol{\rho}_2$, $\Phi_n$ is the index of refraction power spectrum, $z$ is the propagation distance, and $\operatorname{sinc}\left(x\right) = \sin\left(x\right)/x$.  Letting $k_1=k_2=k$ and making the approximation that the second $\operatorname{sinc}\left(x\right)$ function is equal to one (better known as the geometrical optics approximation) simplifies \eq{eq:2} to the expression used for decades to generate atmospheric phase screens.  However, multi-wavelength simulations involving turbulence should use the full expression in \eq{eq:2} to synthesize screens, as it is necessary to capture the physical correlation of the turbulent phase across the wavelengths of interest.           

Charnotskii, in Ref.~\cite{Charnotskii}, was the first to present such a technique.  Therein, he describes how to synthesize two-wavelength phase screens using the sparse spectrum (SS) method~\cite{Charnotskii:13,Charnotskii:13_3}.  Our purpose in this short paper is to show how to generate such screens using the more popular Fourier or spectral method~\cite{Lane, Schmidt:10, Charnotskii:20}.  

It is important to state that although the Fourier technique may be more popular than SS, Charnotskii makes compelling arguments for the superiority of the SS method in Refs.~\cite{Charnotskii:20,Charnotskii}.  We do not dispute his conclusions in those papers.  Rather, our purpose here is merely to generalize the existing single-wavelength Fourier method to any number of wavelengths.  In what follows, we present the theory, procedure, and validation of our multi-wavelength Fourier/spectral phase screen method.

\section{Theory}
We begin with the traditional Fourier method for generating an atmospheric phase screen at a single wavenumber $k_1 = 2\pi/\lambda_1$~\cite{Lane,Schmidt:10,Charnotskii:20}:
\begin{equation}\label{eq:1}
\begin{gathered}
\phi\left[i,j,k_1\right] = \operatorname{Re}\left\{ \sum_{m,n} \left(r^\text{r}\left[m,n,k_1\right] + \text{j}r^\text{i}\left[m,n,k_1\right] \right) \sqrt{\Phi_S\left[m,n,k_1,k_1\right] \frac{2\pi}{L_x}\frac{2\pi}{L_y}} \right. \hfill \\ \left. \quad \times \exp\left(\text{j}2\pi\frac{m}{M}i\right)\exp\left(\text{j}2\pi\frac{n}{N}j\right)\right\}, \hfill
\end{gathered}
\end{equation}
where $\operatorname{Re}\left(c\right)$ is the real part of $c$; $M,N$ are the numbers of points along the horizontal and vertical grid directions; $i,j$ are indices corresponding to $x,y$; and $m,n$ are indices corresponding to spatial frequencies $\kappa_x,\kappa_y$.  Also in \eq{eq:1}, $L_x = M \Delta x$, $L_y = N \Delta y$, $\Delta x, \Delta y$ are the horizontal and vertical grid spacings, $r^\text{r},r^\text{i}$ are $N \times M$ matrices of independent, standard normal random numbers, and lastly, $\Phi_S$ is the phase power spectrum in \eq{eq:2}~\cite{Tatarskii:61,Ishimaru:99}.  Equation~\eqref{eq:1} is equivalent to a discrete inverse Fourier transform; therefore, the fast Fourier transform (FFT) algorithm can be used to synthesize the screen.  Note that the imaginary part of the sum in \eq{eq:1} is statistically independent of the real part and also has the proper spatial statistics.  Either or both can be used to form $\phi$.

Our goal is to generate other phase screens at $k_2,\, k_3,\cdots,\, k_Q$ using \eq{eq:1} that have the proper covariance.  Proceeding, we compute the moment $\Langle \phi\left[i_1,j_1,k_q\right] \phi\left[i_2,j_2,k_p\right] \Rangle$ using \eq{eq:1} and, after expansion, obtain
\begin{equation}\label{eq:3}
\begin{gathered}
\Langle \phi\left[i_1,j_1,k_p\right] \phi\left[i_2,j_2,k_q\right] \Rangle = \sum_{m_1,n_1}\sum_{m_2,n_2} \frac{2\pi}{L_x}\frac{2\pi}{L_y} \sqrt{\Phi_S\left[m_1,n_1,k_p,k_p\right]}\sqrt{\Phi_S\left[m_2,n_2,k_q,k_q\right]} \hfill \\
\quad \times \left\{ \Langle r^\text{r}\left[m_1,n_1,k_p\right] r^\text{r}\left[m_2,n_2,k_q\right] \Rangle \cos\left[2\pi\left(\frac{m_1}{M}i_1 + \frac{n_1}{N}j_1 \right)\right] \cos\left[2\pi\left(\frac{m_2}{M}i_2 + \frac{n_2}{N}j_2\right)\right] \right. \hfill \\ 
\quad \left. +\, \Langle r^\text{i}\left[m_1,n_1,k_p\right] r^\text{i}\left[m_2,n_2,k_q\right] \Rangle \sin\left[2\pi\left(\frac{m_1}{M}i_1 + \frac{n_1}{N}j_1\right)\right] \sin\left[2\pi\left(\frac{m_2}{M}i_2 + \frac{n_2}{N}j_2\right)\right] \right. \hfill \\ 
\quad \left. -\, \Langle r^\text{r}\left[m_1,n_1,k_p\right] r^\text{i}\left[m_2,n_2,k_q\right] \Rangle \cos\left[2\pi\left(\frac{m_1}{M}i_1 + \frac{n_1}{N}j_1\right)\right] \sin\left[2\pi\left(\frac{m_2}{M}i_2 + \frac{n_2}{N}j_2\right)\right] \right. \hfill \\ 
\quad \left. -\, \Langle r^\text{i}\left[m_1,n_1,k_p\right] r^\text{r}\left[m_2,n_2,k_q\right] \Rangle \sin\left[2\pi\left(\frac{m_1}{M}i_1 + \frac{n_1}{N}j_1\right) \right]\cos\left[2\pi\left(\frac{m_2}{M}i_2 + \frac{n_2}{N}j_2\right)\right]\right\}. \hfill
\end{gathered}    
\end{equation}
Letting $\Langle r_1^\text{r} r_2^\text{r} \Rangle = \Langle r_1^\text{i} r_2^\text{i}\Rangle $ and $\Langle r_1^\text{r} r_2^\text{i} \Rangle = \Langle r_1^\text{i} r_2^\text{r} \Rangle$ simplifies \eq{eq:3} to
\begin{equation}\label{eq:4}
\begin{gathered}
\Langle \phi\left[i_1,j_1,k_p\right] \phi\left[i_2,j_2,k_q\right] \Rangle = \sum_{m_1,n_1}\sum_{m_2,n_2} \frac{2\pi}{L_x}\frac{2\pi}{L_y} \sqrt{\Phi_S\left[m_1,n_1,k_p,k_p\right]}\sqrt{\Phi_S\left[m_2,n_2,k_q,k_q\right]} \hfill \\
\quad \times \left\{ \Langle r^\text{r}\left[m_1,n_1,k_p\right] r^\text{r}\left[m_2,n_2,k_q\right] \Rangle \cos\left[2\pi\left(\frac{m_1}{M}i_1 + \frac{n_1}{N}j_1 - \frac{m_2}{M}i_2 - \frac{n_2}{N}j_2\right)\right] \right. \hfill \\ 
\quad \left. -\, \Langle r^\text{r}\left[m_1,n_1,k_p\right] r^\text{i}\left[m_2,n_2,k_q\right] \Rangle \sin\left[2\pi\left(\frac{m_1}{M}i_1 + \frac{n_1}{N}j_1 + \frac{m_2}{M}i_2 + \frac{n_2}{N}j_2\right)\right]\right\}. \hfill
\end{gathered}
\end{equation}
Equation~\eqref{eq:4} must equal $B_S$ in \eq{eq:2} for the screens to be physical.  This requires $\Langle r_1^\text{r} r_2^\text{i} \Rangle = 0$ and
\begin{equation}\label{eq:5}
\begin{gathered}
 \Langle r^\text{r}\left[m_1,n_1,k_p\right] r^\text{r}\left[m_2,n_2,k_q\right] \Rangle = \frac{\Phi_S\left[m_1,n_1,k_p,k_q\right]\delta\left[m_1-m_2\right]\delta\left[n_1-n_2\right]}{\sqrt{\Phi_S\left[m_1,n_1,k_p,k_p\right]}\sqrt{\Phi_S\left[m_2,n_2,k_q,k_q\right]}}, 
 \end{gathered}
\end{equation}
where $\delta\left[x\right]$ is the Kronecker delta function.  Substituting \eq{eq:5} into \eq{eq:4} and evaluating the trivial sums over $m_2,n_2$, we obtain the Riemann sum form of \eq{eq:2}.

The above analysis reveals that we can synthesize phase screens at $k_{p}$ and $k_{q}$ with the proper, physical covariance by using correlated Gaussian random numbers in \eq{eq:1}.  Such numbers are easy to generate using Cholesky factors.  Referring to \eq{eq:5}, the covariance matrix for $r^\text{r}_p, r^\text{r}_q$ and $r^\text{i}_p, r^\text{i}_q$ (the real and imaginary parts are statistically independent) is equal to
\begin{equation}\label{eq:6}
\mathbf{\Sigma}_{pq} = \begin{bmatrix} 1 & R_{pq} \\ R_{pq} & 1\end{bmatrix},
\end{equation}
where $R_{pq}$ is
\begin{equation}
R_{pq} = \frac{\Phi_S\left[m,n,k_p,k_q\right]}{\sqrt{\Phi_S\left[m,n,k_p,k_p\right]}\sqrt{\Phi_S\left[m,n,k_q,k_q\right]}}.    
\end{equation}
It is easy to generalize \eq{eq:6} to any number of wavenumbers, such that
\begin{equation}\label{eq:8}
\mathbf{\Sigma} = \begin{bmatrix} 1 & R_{12} & R_{13} & \dots & R_{1Q} \\ R_{12} & 1 & R_{23} & \dots & R_{2Q} \\  R_{13} & R_{23} & 1 & \dots & R_{3Q} \\ \vdots & \vdots & \vdots & \ddots & \vdots \\
R_{1Q} & R_{2Q} & R_{3Q} & \dots & 1 \end{bmatrix}.
\end{equation}
The Cholesky decomposition of \eq{eq:8}---i.e., $\mathbf{\Sigma} = \mathbf{L}\mathbf{L}^\text{T}$, where $\mathbf{L}$ is a lower triangular matrix---is~\cite{Strang,Watkins}
\begin{equation}\label{eq:9}
\begin{split}
    L_{qq} &= \left(\Sigma_{qq} - \sum\limits_{k=1}^{q-1} L_{qk}^2 \right)^{1/2} \\
    L_{pq} &= \frac{1}{L_{qq}} \left( \Sigma_{pq} - \sum\limits_{k=1}^{q-1} L_{pk}L_{qk}\right) \quad p > q.
\end{split}
\end{equation}
The Gaussian random numbers $r^\text{r}_{1,\cdots,Q}$ and $r^\text{i}_{1,\cdots,Q}$, which are substituted into \eq{eq:1} to synthesize phase screens at $k_1,\, k_2,\, \cdots,\, k_Q$, are generated from $2Q$ $N \times M$ matrices of independent, standard normal random numbers according to
\begin{equation}\label{eq:10}
\begin{bmatrix} r^\text{r}_1, r^\text{i}_1 \\ r^\text{r}_2, r^\text{i}_2 \\ r^\text{r}_3, r^\text{i}_3 \\ \vdots \\ r^\text{r}_Q, r^\text{i}_Q \end{bmatrix} = \begin{bmatrix}
    L_{11} & 0 & 0& \dots & 0 \\
    L_{21} & L_{22} & 0 &\dots & 0 \\
    L_{31} & L_{32} & L_{33} & \dots & 0 \\
    \vdots & \vdots & \vdots & \ddots & \vdots \\
    L_{Q1} & L_{Q2} & L_{Q3} & \cdots & L_{QQ} \end{bmatrix} \begin{bmatrix} r_{1}, r_{Q+1} \\ r_{2}, r_{Q+2} \\ r_{3}, r_{Q+3} \\ \vdots \\ r_{Q}, r_{2Q} \end{bmatrix}
\end{equation}
with $L_{pq}$ computed recursively using \eq{eq:9}.  
\begin{figure}[t]
    \centering
    \includegraphics[width=0.92\textwidth]{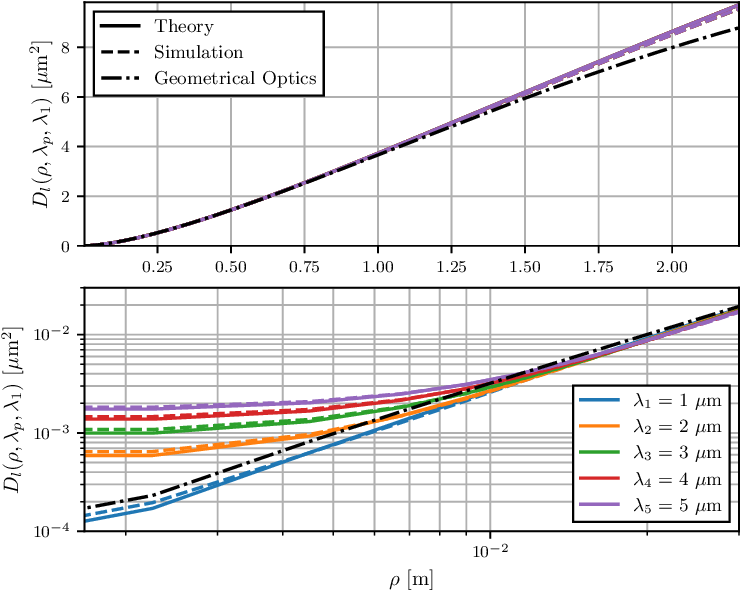}
    \caption{Two-wavelength OPL structure function $D_l\left(\rho,\lambda_p,\lambda_1\right)$ results plotted on linear and log-log scales.}
    \label{fig:SF}
\end{figure}

\section{Procedure}
In summary, the procedure for generating multi-wavelength phase screens using FFTs is as follows:
\begin{enumerate}
    \item Generate two sets of $Q$ standard normal random numbers.
    \item Compute the Cholesky matrix $\mathbf{L}$ using \eq{eq:9}.
    \item Generate $r^\text{r}_{1,\cdots,Q}$ and $r^\text{i}_{1,\cdots,Q}$ using \eq{eq:10}.
    \item Substitute $r^\text{r}_{1,\cdots,Q}$ and $r^\text{i}_{1,\cdots,Q}$ into \eq{eq:1}.
    \item Compute $Q$ two-dimensional FFTs.
\end{enumerate}
We note that our method can also be used to synthesize multi-wavelength subharmonic screens~\cite{Lane,Schmidt:10,Charnotskii:20}:
\begin{enumerate}
    \item Generate two sets of $Q$ standard normal random numbers.
    \item Compute the Cholesky matrix $\mathbf{L}$ using \eq{eq:9}.
    \item Generate $r^\text{r}_{1,\cdots,Q}$ and $r^\text{i}_{1,\cdots,Q}$ using \eq{eq:10}.
    \item Substitute $r^\text{r}_{1,\cdots,Q}$ and $r^\text{i}_{1,\cdots,Q}$ into Eq.~(10) in Ref.~\cite{Charnotskii:20}.
    \item Compute the discrete Fourier transform.
    \item Sum the subharmonic screens.
\end{enumerate}

\section{Validation}
Figure~\ref{fig:SF} shows the optical path length (OPL) structure function $D_l$ results comparing simulation and theory.  The theoretical $D_l$ is given as Eq.~(18) in Ref.~\cite{Charnotskii}, which we computed numerically with $\Phi_n$ equal to the modified von K\'{a}rm\'{a}n spectrum~\cite{Sasiela:07}, $C_n^2 = 3.71 \times 10^{-15} \text{ m}^{-2/3}$, $z = 750 \text{ m}$, $l_0 = 5 \text{ mm}$, and $L_0 = 20 \text{ m}$.  We also included the geometrical optics (GO) $D_l$, which we obtained from the phase structure function $D_S = k_p k_q D_l$ given in Chap.~6, Eq.~(60) of Ref.~\cite{Andrews}.  We computed the simulated $D_l$ from 1,000 statistically independent $\phi$, where the side length and pitch of the square grids were $9 \text{ m}$ and $l_0/3$, respectively.  Each of the 1,000 $\phi$ consisted of an FFT screen plus three subharmonic screens.  We have included a Python script for generating multi-wavelength phase screens in Ref.~\cite{code}.  

We observe excellent agreement of the theoretical and simulated $D_l$ in Fig.~\ref{fig:SF}.  The quality of these results validate our multi-wavelength Fourier phase screen method.  Note that the two-wavelength structure functions do not equal zero at $\rho=0$ like the single-wavelength $D_l$.  This feature is described in Ref.~\cite{Charnotskii}.  In addition, the single-wavelength $D_l$ (blue trace) is below the GO $D_l$ at small separations (bottom plot) due to diffraction, which is included in the phase power spectrum in \eq{eq:1}.  

\section{Conclusion}
In this paper, we presented a method to generate atmospheric phase screens at multiple wavelengths using FFTs.  This work extended the well-established Fourier/spectral method for generating phase screens at a single wavelength and augmented the recent SS method for two-wavelength phase screens~\cite{Charnotskii}.  We began this short paper with a section summarizing the theory underpinning our method and concluded with a step-by-step recipe detailing how to implement our technique.  We then validated our method by comparing the two-wavelength OPL structure function obtained from 1,000 screens at five wavelengths to theory.  The results were in excellent accord.  Our method for generating multi-wavelength phase screens will be useful in simulations studying the effects of strong turbulence on two-wavelength AO systems.

\section*{Acknowledgment(s)}
The authors would like to thank the Joint Directed Energy Transition Office for sponsoring this research.

M.W.H.: The views expressed in this paper are those of the author and do not reflect the policy or position of Epsilon C5I or Epsilon Systems. 

\section*{Disclosure statement}
The author declares that the research was conducted in the absence of any commercial or financial relationships that could be construed as a potential conflict of interest. 

The U.S. Government is authorized to reproduce and distribute reprints for governmental purposes notwithstanding any copyright notation thereon. Distribution Statement A. Approved for public release: distribution is unlimited. Public Affairs release approval \#: NSWCDD-PN-25-00102.

\section*{Funding}
This research received no funding.
 
\bibliographystyle{tfnlm}
\bibliography{main_WRM}

\end{document}